\documentclass[a4paper,oneside,12pt]{article}
\usepackage{amsmath,amssymb,amsfonts}
\usepackage[pdftex]{color,graphicx}

\newcommand{\freccia}[1]{\overrightarrow{#1}}
\newcommand{\p}{\partial}

\begin{document}


\begin{titlepage}
\begin{center}

{\bfseries{\Large Vector-like quarks in a ``composite'' Higgs model}}

\vspace{2.0cm}
{\large Paolo Lodone}

\vspace{2.0cm}
{\itshape Scuola Normale Superiore, Piazza dei Cavalieri 7, I-56126 Pisa, Italy}

\vspace{4cm}

\bfseries{Abstract}

\end{center}

Vector-like quarks are a common feature of ``composite'' Higgs models, where they intervene in cutting off the top-loop contribution to the Higgs boson mass and may, at the same time, affect the Electroweak Precision Tests (EWPT).
A model based on $SO(5)/SO(4)$ is here analyzed. In a specific non minimal version, vector-like quarks of mass as low as 300-500 GeV are allowed in a thin region of its parameter space. Other models fail to be consistent with the EWPT.

\end{titlepage}


\section{Introduction}

The great success of the Standard Model (SM) in predicting the electroweak observables leaves many theoretical open questions. One of them is the famous ``naturalness problem'' of the Fermi scale: one looks for a non-accidental reason that explains why the Higgs boson is so light relatively to any other short distance scale in Physics.

In order to keep the Higgs boson mass near the weak-scale expectation value $v$ with no more than $10 \%$ finetuning it is necessary to cut-off the top, gauge, and scalar loops at a scale $\Lambda_{nat} \lesssim 1-2$ TeV.
This fact tells us that the SM is not natural at the energy of the Large Hadron Collider (LHC), and more specifically new physics that cuts-off the divergent loops has to be expected at or below 2 TeV. In a weakly coupled theory this means new particles with masses below 2 TeV and related to the SM particles by some symmetry.
For concreteness, the dominant contribution comes from the top loop.
Thus naturalness arguments predict new multiplet(s) of top-symmetry-related particles that should be easily produced at the LHC, which has a maximum available energy of 14 TeV.

The possibilities in extending the SM are many. Here we focus on a model (see \cite{contino2}) in which the Higgs particle is realized as a pseudo-Goldstone boson associated to the breaking $SO(5)\rightarrow SO(4)$ at a scale $f > v$. In some sense this extension is ``minimal'' since we add only one field in the scalar sector. The Higgs mass will then be protected from self-coupling corrections, and the cutoff scale can be raised up to $3$ TeV.
Following the approach of \cite{barbieri1}, the $SO(5)$ symmetry has then to be extended to the top sector by adding new vector-like quarks in order to reduce the UV sensitivity of $m_h$ to the top loop. In principle new heavy vectors should also be included in order to cut-off the gauge boson loops, however here only the quark sector will be studied because the dominant contribution comes from the top. Moreover, from a phenomenological point of view, heavy quark searches at the LHC may be easier than heavy vector searches (as pointed out in \cite{barbieri2}).

\begin{figure}[hbt]
\begin{center}
\includegraphics[width=0.6\textwidth]{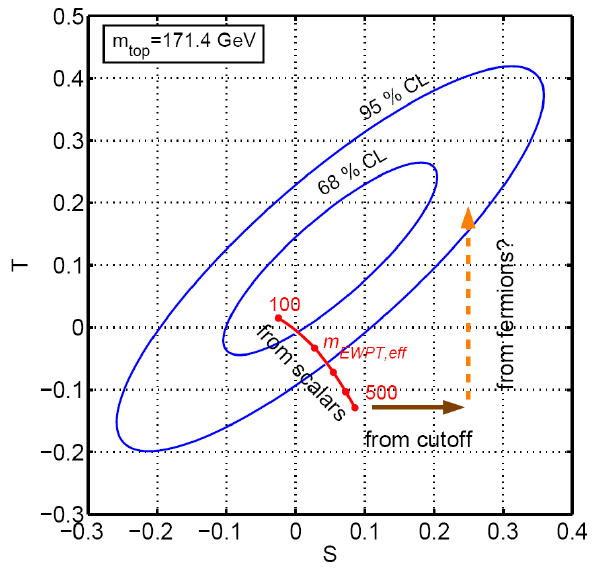}
\end{center}
\caption{The experimentally allowed region in the $ST$ plane, including contributions ``from scalars'' and ``from cutoff'' (see \cite{barbieri1}, section 2). The dashed arrow shows that an extra positive contribution to $T$ is needed in order to make the model consistent with data. In section \ref{extendedmodel} it will be shown that such contribution may come from a suitably extended top sector. This figure is taken from \cite{barbieri1}.}
\label{figuraewpt}
\end{figure}

In enlarging the fermion sector it is necessary to fulfill the requirements of the Electro Weak Precision Tests (EWPT).
More specifically, as shown in Figure \ref{figuraewpt}, the composite nature of the Higgs boson and the physics at the cutoff produce two corrections to the $S$ and $T$ parameters of the SM.
For this reason, in order to be consistent with data, one can look for a positive contribution to $T$ coming from the fermion sector.
Another experimental constraint comes from the modified bottom coupling to the $Z$ boson.

The main virtues of this model are minimality and effectiveness. That is we concentrate on the fermion resonances, which can be lighter than the new gauge bosons and play a central role in reducing the sensitivity of the Higgs boson mass to the new physics. Moreover we do so introducing the least possible number of new particles and parameters. In fact there are models which can be compatible with EWPT data and have the same scalar sector, but since they start from 5d considerations they are forced to introduce much more new fields (see e.g. \cite{contino} and \cite{carena-santiago}).

In section \ref{modelloSO5} a summary of some relevant previous works is reported. In section \ref{extendedmodel} I work out a non minimal model which can be consistent with data. In section \ref{othermodels} two examples are given of other models ruled out by the EWPT.

\section{Summary of previous works} \label{modelloSO5}

Making reference to \cite{contino2} and \cite{barbieri1} for a detailed description of the model, here I concentrate on quarks.
The fermion sector has to be enlarged in such a way that the top is ($SO(5)$ symmetrically) given the right mass $m_t = 171$ GeV, and new heavy quarks are vector-like in the $v/f \rightarrow 0$ limit.
The bottom quark can be considered massless at this level of approximation, while lighter quarks are completely neglected.

The minimal way to do this is to enlarge the left-handed top-bottom doublet $q_L$ to a vector (one for each colour) $\Psi_L$ of $SO(5)$, which under $SU(2)_L \times SU(2)_R$ breaks up as $(2,2)+1$. The SM gauge group $G_{SM}=SU(2)_L \times U(1)$ is here given by the $SU(2)_L$ and the $T_3$ of the $SU(2)_R$ of a fixed subgroup $SO(4)=SU(2)_L \times SU(2)_R \subset SO(5)$.
The full fermionic content of the third quark generation is now:
$$
\Psi_L= \left(
q= \left( \begin{array}{c} t \\ b\end{array}\right) ,\, X= \left( \begin{array}{c} X^{5/3} \\ X \end{array}\right) , \, T
 \right)_L
 \, , \, t_R, X_R= \left( \begin{array}{c} X^{5/3} \\ X  \end{array}\right)_R, T_R ,
$$
where the needed right handed states have been introduced in order to  give mass to the new fermions.
Hypercharges are fixed in order to obtain the correct value of the electric charges.
Note that the upper component of the ``exotic'' $X$ has electric charge $5/3$.

In the next section an extended model with fermions in the fundamental representation will be examined. The spinor representation (see e.g. \cite{contino2}) is ruled out by requiring that the physical left handed b-quark is a true doublet of $SU(2)_L$ and not an admixture of doublet and singlet, as noted in \cite{barbieri1} or in \cite{contino-lett}. The requirement that there be not a left handed charge $-\frac{1}{3}$ singlet to mix whith $b_L$ is a sort of ``custodial symmetry'' which protects the $Zb\overline{b}$ coupling  fom large corrections (\cite{agashe-contino}).

The Yukawa Lagrangian of the fermion sector consists of an $SO(5)$ symmetric mass term for the top (this guarantees the absence of quadratic divergences in the contribution to $m_h$, as shown by equation \ref{diverglogaritm5q}) and the most general (up to redefinitions) gauge invariant mass terms for the heavy $X$ and $T$:
\begin{equation} \label{lagr5iniziale}
\mathcal{L}_{top}= \lambda_1 \overline{\Psi}_L \phi t_R + \lambda_2 f \overline{T}_L T_R + \lambda_3 f \overline{T}_L t_R +M_X \overline{X}_L X_R + h.c,
\end{equation}
where $\phi$ is the scalar 5-plet containing the Higgs Field.
Note that the adjoint representation of $SO(5)$ splits in the adjoint representation of $SO(4)$ plus a $(4)$ of SO(4): this fact guarantees that the Goldstone bosons of the $SO(5)\rightarrow SO(4)$ breaking have the quantum numbers of the Higgs dublet.
Up to rotations that preserve all the quantum numbers, with a convenient definition of the various parameters, we can rewrite \ref{lagr5iniziale} in the form:
\begin{equation} \label{yukawaminimal}
\mathcal{L}_{top}=\overline{q}_L H^c (\lambda_t t_R + \lambda_T T_R) + \overline{X}_L H (\lambda_t t_R + \lambda_T T_R) + M_T \overline{T}_L T_R + M_X \overline{X}_L X_R + h.c.
\end{equation} 

Through diagonalization of the mass matrix we obtain the physical fields, in terms of which it is possible to evaluate the physical quantities.
For example let us check the cancellation of the quadratically divergent contribution to $m_h$ due to the top loop, starting from the potential:
\begin{equation} \label{potenziale}
V= \lambda (\phi^2 -f^2)^2 - A f^2 \freccia{\phi}^2 + B f^3 \phi_5,
\end{equation}
where $\freccia{\phi}$ are the first four components of $\phi$.
The Higgs boson mass can be shown to be controlled by the $A$ parameter, that is by the $SO(5)$-breaking term ($m_h = 2 v \sqrt{A}$ for big $\lambda$). This is reasonable since if everything were $SO(5)$-symmetric the Higgs particle would be a massless Goldstone boson.
The divergent part of the one loop correction to $A$, evaluated as in \cite{barbierisusy} and setting $v=0$ for simplicity, is now:
\begin{eqnarray}
\delta A &=& -\frac{12 f^2}{64 \pi^2} \label{diverglogaritm5q} \lambda_1^2\left(\frac{M_X^2}{f^2}-4\left(\lambda_1+\lambda_3\right)^2-2\lambda_2^2\right)\log\Lambda^2 \nonumber \\
&=& -\frac{3}{16\pi^2 f^2} \left(\lambda_t^2+\lambda_T^2\right) \left(M_X^2 + M_T^2 \left(\frac{2}{1 + \lambda_T^2/\lambda_t^2}-4\right)\right)\log\Lambda^2.
\end{eqnarray}
Notice that there is no quadratic divergence. Moreover $M_X$ and $M_T$ take the role of the cutoff $\Lambda$ in the original top-loop contribution. For this reason we cannot allow them to be much above 2 TeV, otherwise this logarithmic term alone produces a $\delta m_h$ of the same order of the weak-scale expectation value $v$, and we are led again to a naturalness problem.

Some finetuning on the parameters $A, B$ of equation \ref{potenziale} is necessary in order to obtain $v < f$. This can be quantified by the logarithmic derivative:
$$
\Delta = \frac{A}{v^2} \frac{\p v^2}{\p A} \approx \frac{v^2}{f^2}.
$$
To avoid a large $\Delta$, throughout this paper I will assume $f=500$ GeV, which means $\approx 10 \%$ finetune.
This implies that for the ``naturalness cutoff'' of this model we have (see \cite{barbieri1} for a detailed discussion):
$$
\Lambda \approx \frac{4 \pi f}{\sqrt{N_g}} \backsim 3 \mbox{ TeV,}
$$
where $N_g=4$ is the number of Goldstones.

As shown in \cite{barbieri1} this model can be considered as the low energy description of any model in which the EWSB sector has a $SO(5)$ global symmetry partly gauged with $G_{SM}=SU(2)\times U(1)$. Different models can be meaningfully compared at the same level of finetuning, which in practice means the same level of $f$.
Generally, at this level of finetuning, the heavy vector resonances have masses exceeding the cutoff, or at least exceeding the energy scale at which the $WW$-scattering exceeds unitarity in the effective sigma model with the heavy scalar sent above the cutoff\footnote{This is a general consequence of the composite nature of the Higgs Boson, as pointed out in \cite{schmaltz}. For $f=500$ GeV the unitarity is saturated at $s=2.5$ TeV, see \cite{barbieri1}.}.
For this reason it is hard to see any gain in introducing them at all, since they do not substantially improve the calculability. Throughout this paper their contribution is considered to be included in that from the ``physics at the cutoff''.

In order to check the compatibility with the EWPT, it is necessary to evaluate the relative deviations of the $T$ parameter and of the $Z \rightarrow b\overline{b}$ coupling with respect to the usual SM results:
$$
\hat{T}_{SM}=\frac{3 g^2 m_t^2}{64 \pi^2 m_W^2}=\frac{3 G_F m_t^2}{8 \sqrt{2}}.
$$
$$
A^{bb}_{SM}=\frac{\lambda_t^2}{32 \pi^2} \quad , \quad A^{bs}_{SM}=V_{ts}V_{tb}^*A^{bb}_{SM}.
$$
where the definition of $A^{bb}$ and $A^{bs}$ is:
$$
\left( -\frac{1}{2} + \frac{\sin^2 \theta_W}{3} + A^{bb} \right)\frac{g}{\cos \theta_W} Z_\mu \overline{b}_L \gamma^\mu b_L \quad , \quad A^{bs}\frac{g}{\cos \theta_W} Z_\mu \overline{b}_L \gamma^\mu s_L
$$
and the limit $m_t \gg m_W$ is understood (see \cite{barbieri3} or \cite{barbieri4}).
The experimental constraints are summarized in Table \ref{expconstraints}, where the former condition comes from Figure \ref{figuraewpt}, the latter from LEP precision measurements\footnote{See \cite{barbieri1}, par. 3.2.2. Experimental data are from \cite{lepdata}.}.
In principle one could also consider the constraint coming from the b-factories data on $B\rightarrow X_s l^+ l^-$ decays:
$$
\frac{A^{bs}}{A_{SM}^{bs}} = 0.95 \pm 0.20,
$$
however using this constraint or the one in Table \ref{expconstraints}, the final conclusions do not change.

\begin{table}[hbt]
\begin{center}
\begin{tabular}{||c||}
\hline \\
$\quad 0.25 \leq \delta T_{fermions} \leq 0.50 \quad$ \\ \\ \hline \\
$\quad \frac{A^{bb}}{A_{SM}^{bb}} = 0.88 \pm 0.15 \quad$ \\ \\
\hline 
\end{tabular}
\end{center}
\caption{Experimental constraints on $\rho$ and $Z \rightarrow b \overline{b}$.}
\label{expconstraints}
\end{table}

\begin{figure}[phbt]
\begin{center}
\begin{tabular}{cc}
\includegraphics[width=0.46\textwidth]{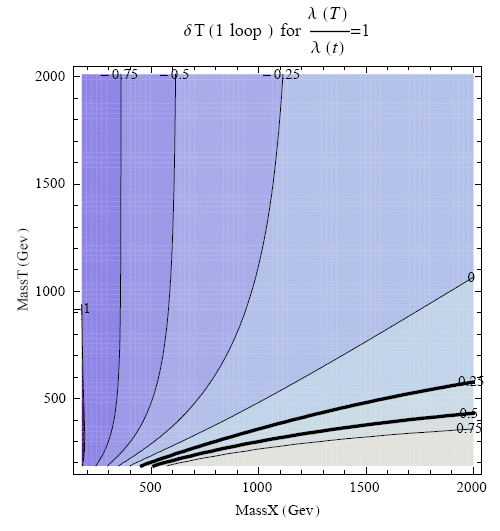} &
\includegraphics[width=0.46\textwidth]{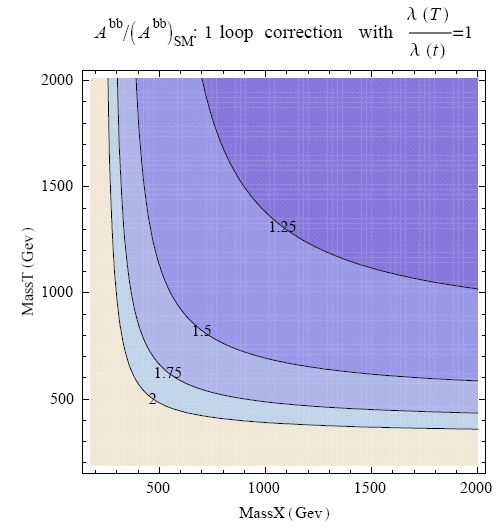}
\end{tabular}
\end{center}
\caption{Numerical isoplot of the one loop corrections to $\delta T$ and $A^{bb}/A^{bb}_{SM}$ versus the Lagrangian parameters $(M_X,M_T)$ in the minimal model for $\lambda_T/\lambda_t= 1$. Note that there is no experimentally allowed region.}
\label{plot5q1.0}
\end{figure}

Analytic approximate expressions for $\delta T$ and $\delta A^{bb}$ can be found in \cite{barbieri1}.
In Figure \ref{plot5q1.0} a typical result of the numerical computation of the one-loop $\delta T$ and $A^{bb}/A^{bb}_{SM}$ is reported in terms of the parameters of Lagrangian \ref{yukawaminimal}.
The only effective free parameters are $M_X$, $M_T$ (which are roughly equal to the physical masses) and $\lambda_t / \lambda_T$, which is taken from 1/3 to 3 so that the theory is not strongly coupled.
The result is that there are no allowed regions in the parameter space for this minimal model. This fact suggests to consider the non minimal model of the next section.

\section{An extended model in the top sector} \label{extendedmodel}

In this section I study an explicit extended top sector which is motivated only by the fulfillment of the requirements of the EWPT.

\subsection{Non-minimal model} \label{fenomenologia7q}

In this non minimal model the $SO(5)$ symmetric quark sector is completely made up of new quarks and the top mass term arises through order $v/f$ mass mixing. The fermionic content is now:
$$
\Psi_{L,R}= \left(
Q= \left( \begin{array}{c} Q^u \\ Q^d \end{array}\right) ,\, X= \left( \begin{array}{c} X^u \\ X^d \end{array}\right) , \, T
 \right)_{L,R}
 \, , \, q_L= \left( \begin{array}{c} t \\ b \end{array}\right)_L \, , \, t_R,
$$
where $Q$ is now a standard ($Y=1/6$) $SU(2)_L$ doublet and the quantum numbers are the same as in the previous case.
The Yukawa Lagrangian is now $\mathcal{L} = \mathcal{L}_{int}+\mathcal{L}_{BSM}$, where $\mathcal{L}_{BSM}$ involves only ``beyond the SM'' fields with a non renormalizable Yukawa interaction, and $\mathcal{L}_{int}$ describes the mass mixing of the standard fields with the heavy fermions:
\begin{eqnarray} \label{interazioneNONRIN}
\mathcal{L}_{BSM} &=& \frac{y}{f} \overline{\Psi}_L \phi \phi^T \Psi_R + m_Q \overline{Q}_L Q_R + m_X \overline{X}_L X_R+ m_T \overline{T}_L T_R +h.c. \nonumber \\
\mathcal{L}_{int} &=&  \lambda_1 f \overline{q}_{L}Q_R + \lambda_2 f \overline{T}_L t_R +h.c. \end{eqnarray}
Defining:
$$
\lambda_t= \frac{y \lambda_1 \lambda_2 f^2}{\sqrt{(m_T+y f)^2+(\lambda_2 f)^2} \sqrt{m_Q^2+(\lambda_1 f)^2}}
$$
$$
\begin{array}{ll}
M_T = \sqrt{(m_T+y f)^2+(\lambda_2 f)^2} & A=\frac{m_T+y f}{\lambda_2 f}\\
M_Q = \sqrt{m_Q^2+(\lambda_1 f)^2} & B=\frac{m_Q}{\lambda_1 f} \\
M_X =m_X & M_f=\lambda_t f \qquad,
\end{array}
$$
up to rotations which preserve quantum numbers the charge $2/3$ mass matrix becomes, with an obvious notation and ``quark vectors'' $(t,T,Q,X)_{L,R}$:
\begin{equation} \label{massmatrix7q}
\left(\begin{array}{cccc}
\lambda_t v & - A \lambda_t v & -\frac{\sqrt{1+A^2}(\lambda_t v)^2}{M_f} & -\frac{\sqrt{1+A^2}(\lambda_t v)^2}{M_f} \\
0 & M_T & \sqrt{1+A^2}\sqrt{1+B^2} \lambda_t v & \sqrt{1+A^2}\sqrt{1+B^2} \lambda_t v \\
- B \lambda_t v & A B \lambda_t v & M_Q + \frac{B \sqrt{1+A^2}(\lambda_t v)^2}{M_f} & \frac{B \sqrt{1+A^2}(\lambda_t v)^2}{M_f} \\
- \sqrt{1+B^2} \lambda_t v & A \sqrt{1+B^2} \lambda_t v & \frac{ \sqrt{1+A^2}\sqrt{1+B^2}(\lambda_t v)^2}{M_f} & M_X +\frac{ \sqrt{1+A^2}\sqrt{1+B^2}(\lambda_t v)^2}{M_f}
\end{array}\right) .
\end{equation}
The physical masses of the charge $2/3$ quarks will be corrected by diagonalization, while the $Q^d$ (charge $-\frac{1}{3}$) and $X^u$ (charge $\frac{5}{3}$) masses remain exactly $M_Q$ and $M_X$ since there is no state to mix with.
As already mentioned, to avoid finetuning we shall take $f=500$ GeV so that $M_f$ is not a free parameter.

I report the exact one loop results for $\delta T$ and $A^{bb}$ up to order $\epsilon^2$ in the limit in which three masses are much bigger than the other one.

For the correction to $T$ we have:
\begin{eqnarray} \label{deltaT7qNonrinOrd2}
M_Q,M_X,M_f>>M_T:&  & \frac{\delta T}{T_{SM}}\approx   2A^2(\log\frac{M_T^2}{m_t^2} - 1 + \frac{A^2}{2})(\frac{m_t}{M_T})^2 \nonumber \\
M_T,M_X,M_f>>M_Q: &  & \frac{\delta T}{T_{SM}}\approx   4B^2(\log\frac{M_Q^2}{m_t^2} - \frac{3}{2} + \frac{1}{3}B^2)(\frac{m_t}{M_Q})^2 \\
M_T,M_Q,M_f>>M_X: &  & \frac{\delta T}{T_{SM}}\approx   -4(1+B^2)(\log\frac{M_X^2}{m_t^2} -\frac{11}{6}-\frac{1}{3} B^2 )(\frac{m_t}{M_X})^2 \nonumber
\end{eqnarray}
$M_T,M_Q,M_X>>M_f$:
\begin{eqnarray*}
\frac{\delta T}{T_{SM}} &\approx &  \frac{2(1+A^2)}{3(M_Q^2-M_X^2)^2} \{ 12 B \sqrt{1+B^2}(M_Q^3 M_X+M_Q M_X^3) - \\ &&- (1+2B^2)(7(M_Q^4+M_X^4)-26M_Q^2 M_X^2) + \\&& + \frac{6\log\frac{M_Q^2}{M_X^2}}{M_Q^2-M_X^2}(-4B\sqrt{1+B^2}M_Q^3 M_X^3-3M_Q^2 M_X^4 + M_X^6 + \\ && +B^2(M_Q^6 -3M_Q^4M_X^2-3M_Q^2M_X^4+M_X^6)) \}(\frac{\lambda_t v}{M_f})^2 .
\end{eqnarray*}
while for $Z \rightarrow b \overline{b}$ it is:
\begin{eqnarray}
M_Q, M_X>>M_T: & & \quad \frac{\delta A^{bb}}{A_{SM}^{bb}} \approx  2A^2(\log\frac{M_T^2}{m_t^2}-1+\frac{A^2}{2})(\frac{m_t}{M_T})^2 \label{Zinbb7qNonrinOrd2} \\
M_T, M_X>>M_Q: & & \frac{\delta A^{bb}}{A_{SM}^{bb}} \approx  B^2(\log\frac{M_Q^2}{m_t^2}-1)(\frac{m_t}{M_Q})^2 + 2B \sqrt{1+A^2} \frac{(\lambda_t v)^2}{M_Q M_f} \nonumber
\end{eqnarray}
$M_T, M_Q>>M_X:$
$$
\frac{\delta A^{bb}}{A_{SM}^{bb}} \approx  (1+B^2)(\log\frac{M_X^2}{m_t^2}-1)(\frac{m_t}{M_X})^2 + 2\sqrt{1+B^2} \sqrt{1+A^2} \frac{(\lambda_t v)^2}{M_X M_f}
$$
These results are compatible with \cite{carena-santiago}.
In the following, through numerical diagonalization of the mass matrix, it will be shown that compatibility with the experimental constraints of Table \ref{expconstraints} is now allowed in a thin slice of parameter space.

\subsection{Minimal values for the masses of the new quarks} \label{minimalmasses}

\begin{figure}[phbt]
\begin{center}
\begin{tabular}{cc}
\includegraphics[width=0.46\textwidth]{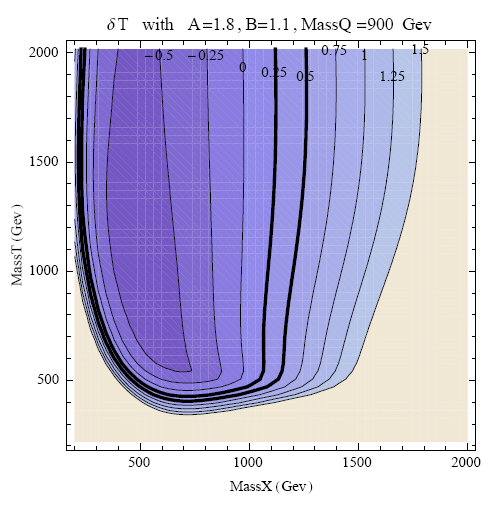} &
\includegraphics[width=0.46\textwidth]{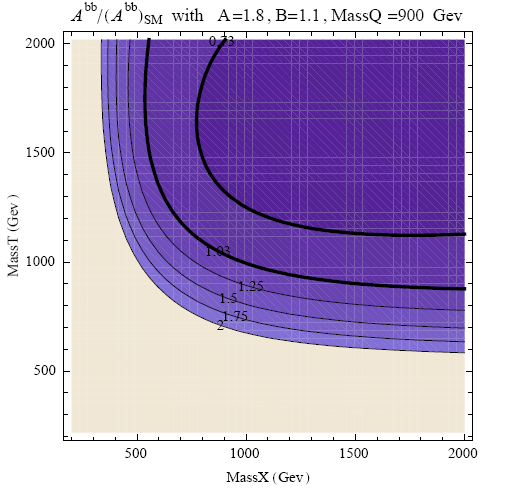}
\end{tabular}
\end{center}
\caption{Isoplot of $\delta T$ and $A^{bb}/A^{bb}_{SM}$ in the $(M_X,M_T)$ plane for ($A=1.8$, $B=1.1$, $M_Q=900$ GeV) in the non-minimal model.}
\label{plot1.8_1.1_Q900}
\end{figure}

The parameter space has been studied for $\frac{1}{3}\leq A,B\leq 3$ with vector-like quark masses all below $M_{max}$, looking for experimentally allowed configurations with relatively light vector-like quarks.
For naturalness considerations, $M_{max}$ cannot be much above 2 TeV (see equation \ref{diverglogaritm5q}).
A typical situation, for example $A=1.8$, $B=1.1$, $M_Q=900$ GeV, is represented in Figure \ref{plot1.8_1.1_Q900}, where I report the isolines of $\delta T$ and $A^{bb}$ in the $(M_X,M_T)$ plane. The thicker lines correspond to the regions constrained as in Table \ref{expconstraints}. The small overlap between the two regions around $M_X, M_T \approx$ 1 TeV is the allowed portion of the parameter space.

For a better illustration of this case, consider for example an exact one loop calculation, which corresponds to a point in Figure \ref{plot1.8_1.1_Q900} (masses in GeV):
\begin{center}
\begin{tabular}{lcr}
\begin{tabular}{|ll|}
\hline
$M_T=1000$ & $M_X=1100$ \\
$M_Q=900$ &  \\
$A=1.8$ & $B=1.1$ \\
\hline
\end{tabular} & $ \rightarrow $ &
\begin{tabular}{|ll|}
\hline
$M_T^{phys}=1010$ &  \\
$M_Q^{phys}=550$ & $M_{Q^{-1/3}}^{phys}=920$\\
$M_X^{phys}=1940$ & $M_{X^{5/3}}^{phys}=1130$ \\
\hline
$\delta T = 0.28$ & $A^{bb}= 0.97$ \\
\hline
\end{tabular}
\end{tabular}
\end{center}
Note the significant difference between $M_{heavy}$ and $M_{heavy}^{phys}$. This is mainly due to diagonalization splitting, but also to a mass-matrix rescaling which is necessary in order to get the correct value for the top mass.

A most interesting phenomenological question concerns the smallest possible values for the masses of the new quarks which are compatible with the constraints of Table \ref{expconstraints}. A study of the full parameter space allows to assert that the following properties hold:
\begin{enumerate}
\item At least one of the new charge 2/3 quarks has to be heavy, that is around 1.9 TeV.
\item \emph{Light Q:} The lightest possible new-quark state is the $Q^{2/3}$, which in principle can be as light as 290 GeV. In such configurations a heavy $T$ or $X$ is required, for example:
\begin{center}
\begin{tabular}{lcr}
\begin{tabular}{|ll|}
\hline
$M_T=1225$ & $M_X=630$ \\
$M_Q=320$ &  \\
$A=2.81$ & $B=0.33$ \\
\hline
\end{tabular} & $ \rightarrow $ &
\begin{tabular}{|ll|}
\hline
$M_T^{phys}=1890$ &  \\
$M_Q^{phys}=290$ & $M_{Q^{-1/3}}^{phys}=340$\\
$M_X^{phys}=555$ & $M_{X^{5/3}}^{phys}=670$ \\
\hline
$\delta T = 0.40$ & $A^{bb}= 1.02$ \\
\hline
\end{tabular}
\end{tabular}
\end{center}
\item \emph{Light T:} Allowing $M_X \approx$ 1.9 TeV it is possible to obtain a $T$ quark mass around $500$ GeV for example:
\begin{center}
\begin{tabular}{lcr}
\begin{tabular}{|ll|}
\hline
$M_T=940$ & $M_X=1200$ \\
$M_Q=960$ &  \\
$A=1.86$ & $B=1.1$ \\
\hline
\end{tabular} & $ \rightarrow $ &
\begin{tabular}{|ll|}
\hline
$M_T^{phys}=510$ &  \\
$M_Q^{phys}=1060$ & $M_{Q^{-1/3}}^{phys}=955$\\
$M_X^{phys}=1940$ & $M_{X^{5/3}}^{phys}=1195$ \\
\hline
$\delta T = 0.43$ & $A^{bb}= 1.01$ \\
\hline
\end{tabular}
\end{tabular}
\end{center}
\item \emph{Light X:} $M_{X^{2/3}}$ can be as low as 450 GeV with $X^{5/3}$ at 950 GeV, and heavy $T$, for example:
\begin{center}
\begin{tabular}{lcr}
\begin{tabular}{|ll|}
\hline
$M_T=1152$ & $M_X=969$ \\
$M_Q=971$ &  \\
$A=2.99$ & $B=0.71$ \\
\hline
\end{tabular} & $ \rightarrow $ &
\begin{tabular}{|ll|}
\hline
$M_T^{phys}=2050$ & \\
$M_Q^{phys}=925$ &$M_{Q^{-1/3}}^{phys}=1026$\\
$M_X^{phys}=460$ & $M_{X^{5/3}}^{phys}=1024$ \\
\hline
$\delta T = 0.28$ & $A^{bb}= 0.93$ \\
\hline
\end{tabular}
\end{tabular}
\end{center}
\item \emph{Light $X^{5/3}$:} $M_{X^{5/3}}$ can be relatively small. From point 2 we see that the $X^{5/3}$ can be also as light as 670 GeV.
\end{enumerate}

\subsection{Allowed volume in parameter space} \label{parameterspace}

\begin{figure}[phbt]
\begin{center}
\includegraphics[width=0.75\textwidth]{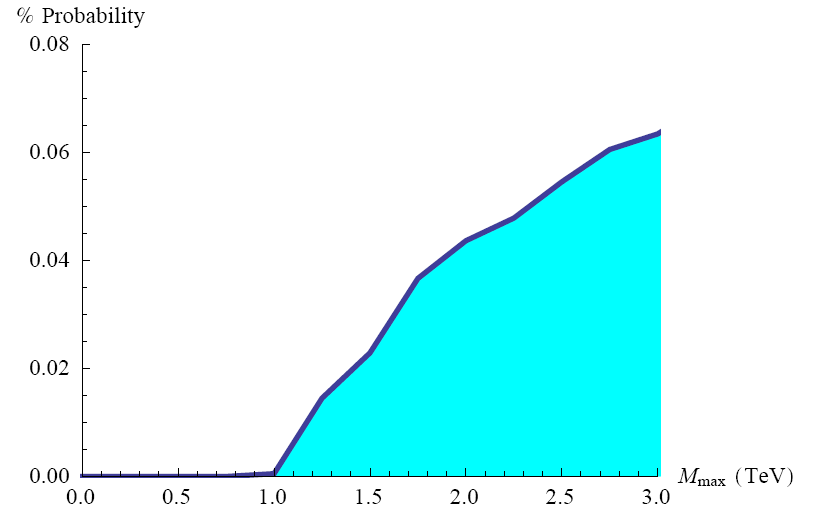}
\end{center}
\caption{\% Probability of the allowed region (see text).}
\label{spazioparametri}
\end{figure}

Of some interest is the following question: how extended is the volume of the parameter space which is allowed by the experimental data?
To answer this question one can consider the fractional volume (making a linear sampling) of the experimentally allowed region in the relevant parameter space:

$$
\left\{ \frac{1}{3} \, \leq \, A, B \, \leq \, 3 \right\} \cap \left\{ 200 \mbox{ GeV } \leq \, M_{T, X, Q} \, \leq M_{max} \right\}
$$
I call ``probability'' of the model this fractional volume.
Note that all the points in the ``total volume'' of this parameter space are viable in the sense of giving a correct EWSB, even if most of them do not satisfy the EWPT.
In Figure \ref{spazioparametri} the result of this calculation is given as a function of $M_{max}$.
For example we have:
\begin{equation}\label{probability}
\frac{\mbox{Allowed volume}}{\mbox{Total volume }(M_{max} = 2.5 \mbox{ TeV})}\approx 0.05 \% = \frac{1}{2000} \quad .
\end{equation}
Note that for the model to be consistent with data it is necessary to have at least one $M_{heavy} \gtrsim 1$ TeV (which actually leads to one $M_{heavy}^{phys} \gtrsim 1.8$ TeV because of the mass splitting and rescaling, as explained in section \ref{minimalmasses}).

\section{Alternative models} \label{othermodels}

In the course of this investigation other Lagrangian models for fermion masses have been considered, all involving more fields than the minimal model. In this section I briefly report about two of them, with different motivations. None gives an acceptable region of parameter space.

\subsection{A different coupling} \label{so5rinormalizz}

One can ask himself what happens considering Lagrangian \ref{interazioneNONRIN} with a standard Yukawa fermion-scalar interaction instead of the non-renormalizable one studied in the previous section.

This can be done by taking exactly the same fermion sector extension described in section \ref{fenomenologia7q} with:
\begin{eqnarray} \label{interazioneRIN}
\mathcal{L} &=&  \lambda_1 f \overline{q}_{L}Q_R + \lambda_2 f \overline{T}_L t_R +y \overline{\Psi}_L \phi t_R\nonumber \\
 &&  + m_Q \overline{Q}_L Q_R + m_X \overline{X}_L X_R+ m_T \overline{T}_L T_R +h.c. \end{eqnarray}
Note that in this case there is no separation between $\mathcal{L}_{int}$ and $\mathcal{L}_{BSM}$ as in the model of section \ref{extendedmodel}.
Up to rotations which preserve quantum numbers the charge $2/3$ mass matrix is now, with the same notation of \ref{massmatrix7q}:
\begin{equation} \label{massmatrix7qRIN}
\left(\begin{array}{c} \overline{t}_L^0 \\ \overline{T}_L^0 \\ \overline{Q}_L^0 \\ \overline{X}_L^0 \end{array}\right)
\left(\begin{array}{cccc}
-\lambda_t v & - A \lambda_t v & 0 & 0 \\
0 & M_T & 0 & 0 \\
B \lambda_t v & A B \lambda_t v & M_Q & 0 \\
- \sqrt{1+B^2} \lambda_t v & A \sqrt{1+B^2} \lambda_t v & 0 & M_X
\end{array}\right)
\left(\begin{array}{c} t_R^0 \\ T_R^0 \\ Q_R^0 \\ X_R^0 \end{array}\right) +h.c,
\end{equation}
where it is:
$$
\lambda_t= \frac{y \lambda_1 f^2}{\sqrt{1+\frac{f^2(\lambda_2 +y )^2}{(m_T)^2}} \sqrt{(m_Q+(\lambda_1 f)^2}}
$$
$$
\begin{array}{ll}
M_T = \sqrt{m_T^2+f^2(\lambda_2 +y)^2} & A=\frac{(\lambda_2+y) f}{m_T}\\
M_Q = \sqrt{m_Q^2+(\lambda_1 f)^2} & B=\frac{m_Q}{\lambda_1 f} \\
M_X =m_X .  & 
\end{array}
$$
Repeating the procedure to evaluate the $\delta T$ and $A^{bb}$ corrections, we obtain exactly the expressions \ref{deltaT7qNonrinOrd2} and \ref{Zinbb7qNonrinOrd2} with $M_f \rightarrow \infty$.
Note however that the definition of $A$  and $B$ is different.

The result of the numerical calculation is that there is no experimentally allowed region. This fact shows also that compatibility with the EWPT is a delicate issue, and in general the approximation $\frac{m_t}{M_{heavy}}\ll 1$ is not reliable.

\subsection{A different model: $SU(4)/Sp(4)$}

In the literature several other models for the Higgs particle as a pseudo-Goldstone bosone have been considered, based on different groups. A common feature shared with the $SO(5)/SO(4)$ model is an extended top sector. Here we consider a suitable extension of the $SU(4)/Sp(4)$ composite-Higgs theory described in \cite{katz}.
The number of fields which can mix with the top is exactly the same as in the $SO(5)/SO(4)$ extended model, with the same number of free parameters.

Enlarging the top sector with new fermions in the vectorial representation of $SU(4)$, as done in \cite{katz}, is problematic because there is an $SU(2)_L$ left handed singlet with electric charge $-1/3$ which will mix at tree level with the bottom, and this is phenomenologically not defendable (see section \ref{modelloSO5}).
This problem is avoided with new fermions in the antisymmetric representation:
$$
A_L=\left(
\begin{array}{cccc}
0 & Q_L & X_L^{5/3} & t_L \\
 & 0 & X_L & b_L \\
 & & 0 & T_L \\
 & & & 0
\end{array}
\right).
$$
The quantum numbers of the fields are fixed by the natural way $SU(2)_L \times SU(2)_R$ is embedded in $Sp(4)$. Introducing the needed right-handed states the third generation is therefore enlarged as:
$$
\left(\begin{array}{c}X^{5/3}_{L,R} \\ X_{L,R} \end{array}\right)=(2)_{7/6} \quad , \quad \left(\begin{array}{c}t_L \\ b_L \end{array}\right)=(2)_{1/6} \quad , \quad t_R, Q_{L,R}, T_{L,R}=(1)_{2/3} .
$$

The most general Lagrangian respecting $SU(2)\times U(1)$ gauge invariance and the $SU(4)$ symmetry of the Yukawa interaction is:
\begin{eqnarray*}
\mathcal{L} &=& \lambda_1 f \overline{t}_R Q_L + \lambda_2 f \overline{t}_R T_L + \frac{1}{2} y_1 \overline{Q}_R tr(\Sigma^* A_L) + y_2 f \overline{Q}_R T_L \\
&& + m_Q \overline{Q}_R Q_L + m_T \overline{T}_R T_L + m_X \left( \overline{X}_R X_L + \overline{X}^{5/3}_R X^{5/3}_L \right)
\end{eqnarray*}
where, keeping only the Yukawa interactions with the Higgs doublet:
$$
\frac{1}{2} tr (\Sigma^* A_L)= f (Q_L + T_L) + H \left(\begin{array}{c}t_L \\ b_L \end{array}\right) + H^c \left(\begin{array}{c}X^{5/3}_L \\ X_L \end{array}\right).
$$
Here $f$ is the scale of the $SU(4)/Sp(4)$ breaking and $H$ is the Higgs doublet.

This Lagrangian can be analyzed in a totally analogous way as in the previous sections.
The mass matrix, concentrating on charge $2/3$ quark mass terms and up to quantum-number preserving rotations, is:
\begin{equation} \label{massmatrix}
\mathcal{L}= \left(\begin{array}{c}
\overline{t}_L \\ \overline{T}_L \\ \overline{Q}_L \\ \overline{X}_L^d \end{array}\right)^T
\left(\begin{array}{cccc}
\lambda_t v & A \lambda_t v & B \lambda_t v & 0 \\
0 & M_T & 0 & 0 \\
0 & 0 & M_Q & 0\\
\lambda_t v & A \lambda_t v & B \lambda_t v & M_X
\end{array}\right)
\left(\begin{array}{c} t_R \\ T_R \\ Q_R \\ X_R^d \end{array}\right) + h.c.
\end{equation}
where for example:
$$
\lambda_t= \frac{\lambda_1 y_1 m_T f}{\sqrt{2}\sqrt{f^2 \lambda_1^2 + (m_Q + f y_1)^2}\sqrt{m_T^2+\frac{f^2(\lambda_2(m_Q+f y_1)-f \lambda_1 (y_1+y_2))^2}{f^2 \lambda_1^2+(m_Q+f \lambda_1)^2}}},
$$
and also the other new parameters are combinations of the original ones.
Note that now $Q$ is a singlet like $T$, while $X$ is again a component of an $Y=7/6$-doublet.
Computing the one loop correction to the $T$ parameter up to second order in $\lambda_t v /M_{heavy}$ we now obtain:
\begin{eqnarray*}
M_X,M_Q>>M_T: && \frac{\delta T}{T_{SM}}\approx   2 A^2 (\log\frac{M_T^2}{m_t^2} - 1 + \frac{A^2}{2})(\frac{m_t}{M_T})^2 \\
M_X,M_T>>M_Q: && \frac{\delta T}{T_{SM}}\approx   2B^2(\log\frac{M_Q^2}{m_t^2} - 1 + \frac{B^2}{2})(\frac{m_t}{M_Q})^2 \\
M_Q,M_T>>M_X: && \frac{\delta T}{T_{SM}}\approx   -4(\log\frac{M_X^2}{m_t^2} -\frac{11}{6} )(\frac{m_t}{M_X})^2
\end{eqnarray*}
while for $Z \rightarrow b \overline{b}$ it is:
\begin{eqnarray*}
M_Q, M_X>>M_T: && \frac{\delta A^{bb}}{A_{SM}^{bb}} \approx  2A^2(\log\frac{M_T^2}{m_t^2}-1+\frac{A^2}{2})(\frac{m_t}{M_T})^2 \\
M_T, M_X>>M_Q: && \frac{\delta A^{bb}}{A_{SM}^{bb}} \approx  2B^2(\log\frac{M_Q^2}{m_t^2}-1+\frac{B^2}{2})(\frac{m_t}{M_Q})^2 \\
M_Q, M_T>>M_X: && \frac{\delta A^{bb}}{A_{SM}^{bb}} \approx  (\log\frac{M_X^2}{m_t^2}-\frac{1}{2})(\frac{m_t}{M_X})^2
\end{eqnarray*}

\begin{figure}[phbt]
\begin{center}
\begin{tabular}{cc}
\includegraphics[width=0.46\textwidth]{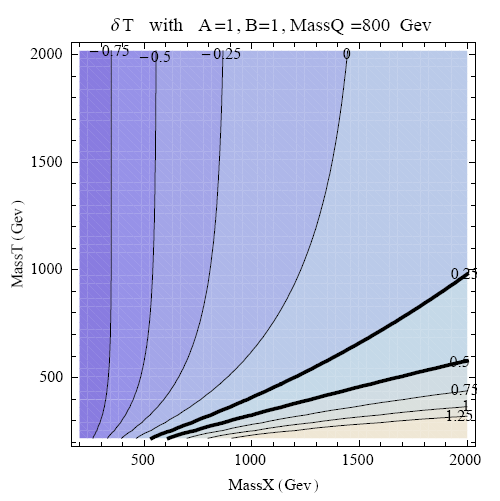} &
\includegraphics[width=0.46\textwidth]{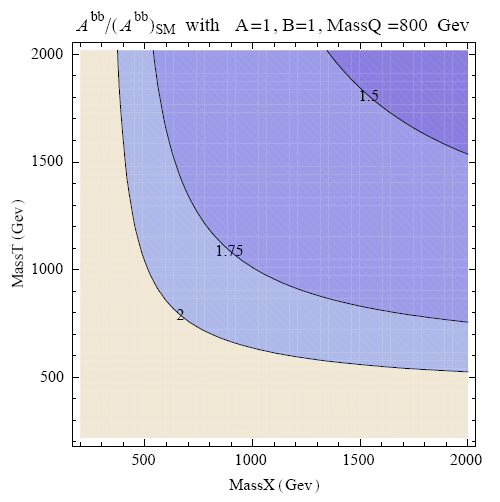}
\end{tabular}
\end{center}
\caption{Isolines of $\delta T$ and $A^{bb}/A^{bb}_{SM}$ in the $(M_X,M_T)$ plane for the the $SU(4)/Sp(4)$ model. Plot for $A=B=1, \, M_Q=800$ GeV.}
\label{plotsu4sp4}
\end{figure}

The experimental consistency of the model has been checked via numerical diagonalization of the mass matrix (\ref{massmatrix}) in the relevant parameter space: $\frac{1}{3} \leq A,B \leq 3$ and $M_T, M_Q, M_X$ below 2 TeV.
The final result is that this model can not be consistent with experimental data. In Figure \ref{plotsu4sp4} I give an example of the typical situation.

\section{Conclusions}

Heavy vector-like fermions are a likely component of models for Electroweak symmetry breaking which address the naturaleness problem of the Fermi scale.
Constraining their mass is crucial in order to assess the potential of their discovery at the LHC.
Here I have analyzed such constraints in a $SO(5)/SO(4)$ model for the Higgs doublet as a pseudo-Goldstone boson. These constraints arise from the EWPT, including B-physics.

Confirming the results of \cite{barbieri1}, I have found that the minimal extension of the top sector has problems in fulfilling the experimental requirements.
For this reason I have considered other possible extensions of the fermion sector, as well as another model based on a different symmetry. These models have received attention in the literature and have different motivations.

The main result is that one such extension is consistent with the constraints coming from the EWPT, including B-physics, in a thin region 
of its parameter space.
To the third generation quarks of the Standard Model one has to add a full vector-like 5-plet of SO(5), i.e. in particular three new quarks of charge 2/3 which mix with the top: $T,Q,X$.

In this region of parameter space the new quarks can be as light as a few hundreds GeV and might therefore be accessible at the LHC.
The range of possible masses is summarized in the following Table (see section \ref{minimalmasses}):

\begin{center}
\begin{tabular}{|c|c|c|} \hline
Quark & $SU(2)_L \times U(1)_Y$ & Constraints on mass  \\ \hline \hline
$Q$ & $(2)_{1/6}$ & $M_{Q^{2/3}} \gtrsim 300$ GeV, $M_{Q^{-1/3}} \gtrsim 350$ GeV \\ \hline
$T$ & $(1)_{2/3}$ & $M_{T^{2/3}} \gtrsim 500$ GeV \\ \hline
$X$ & $(2)_{7/6}$ & $M_{X^{2/3}} \geq$ 450 GeV, $M_{X^{5/3}} \gtrsim$ 650 GeV \\ \hline
\end{tabular}
\end{center}

It is of interest that, randomly ``picking up a point'' in the relevant parameter space with all fermion masses below 2.5 TeV, the probability of being consistent with data is very small, roughly $1/2000$.

None of the other similar models that have been examined have regions of the corresponding parameter space which are compatible with the experimental data.

\section*{Acknowledgements}

For this work I am greatly indebted to Riccardo Barbieri. I also thank Vyacheslav S. Rychkov, Duccio Pappadopulo, and Giovanni Pizzi for useful discussions.



\end{document}